# CROSS-CONTEXTUAL USE OF INTEGRATED INFORMATION SYSTEMS

Gasparas, Jarulaitis, Norwegian University of Science and Technology, Sem Sælands vei 7-9, 7491 Trondheim, Norway, Gasparas@idi.ntnu.no

Eric, Monteiro, Norwegian University of Science and Technology, Sem Sælands vei 7-9, 7491 Trondheim, Norway, Eric.Monteiro@idi.ntnu.no

## Abstract

*Large-scale organizations are increasingly promoting more collaborative and collective work practices across organizational boarders. A predominant way to achieve better collaboration in large-scale heterogeneous contexts is to establish an integrated and standardized technological infrastructure. Ethnographically inspired studies, on the other hand, have challenged such perspective and illustrated that generic technology does not fit in local contexts and needs to be worked-around. Similarly, this paper empirically exemplifies local workarounds and illustrates ongoing and persistently imperfect integration of a collaborative infrastructure in a global oil and gas company. More importantly, however, our analysis focuses on how integrated technology is used across contexts. We illustrate how local workarounds, as a result of tight technological integration, shape use patterns across contexts. Integrated systems establish interdependencies across contexts, thus, the use implies cross-contextual rather than local enactment. Since the trajectory of enactment is influenced by cross-contextual constrains, our study is addressing the existing overemphasis on studying/analysing the use of technology in isolated local contexts. Practically, our study suggests considering workarounds as an intrinsic part of every day work, which should be calculated as additional costs of making the generic technology to work in practice.*

# 1 INTRODUCTION

Abandoning earlier and overly structuralist accounts, there has been a steady increase in information systems research exploring contextual aspects of how technology is developed and used (Avgerou and Ciborra 2004). Ethnographically inspired studies have demonstrated beyond any reasonable doubt the situated nature of how information systems are appropriated (Orlikowski 1996; Robey and Sahay 1996; Walsham 2001). Theoretically, there has been an 'agentic turn', which has "led increasingly to the theoretical positions that privilege human agency over social structures and technological futures" (Boudreau and Robey 2005, p.3), for instance by advocating how technology is always 'enacted' (Orlikowski 2000). The locus of attention is local work practices and how technology is enacted in a situated context, where context is limited to individual actor's engagement and "recurrent interaction with the technology at hand" (Orlikowski 2000, p. 47). Since, enactments more than often deviate from intended system design, a practical concern, then, relates to whether workarounds need to be eliminated (Azad and King 2008) or considered as an intrinsic part of every day work (Rolland and Monteiro 2002).

Integrated, collaborative systems (e.g. enterprise resource planning (ERP) systems, coordination technology (Lotus Notes, MS SharePoint), customer relationship management (CRM) systems) are attractive to business and public sector for their promise to promote more collaborative and collective work practices. Working more collectively across geographical, professional and organizational boundaries entails that one previously local, independent context of use gets linked with (i.e. becomes dependent on) other contexts. As opposed to largely local, independent contexts of enacted technology, use of integrated systems implies the *interdependent* enactment *across* the contexts now linked as a result of the *integration*.

The main purpose of this paper is to analyse the form and implications of cross-contextual enactment of integrated systems. We explore questions such as: how does one local workaround affect other contexts of use; how does local appropriation of technology 'travel' to other contexts mediated by the integration, possibly creating unintended consequences there?

The empirical basis of our paper is an ongoing, longitudinal (2007-2008) case study of a global oil and gas company (OGC, a pseudonym to maintain anonymity) where we also earlier (1997-1998) studied integrated, collaborative systems (Monteiro and Hepsø 2002). Operating across significant geographical, professional, business and organizational boundaries, OGC is struggling to move towards more collaborative modes of working. Integrated systems are a strategically recognised vehicle to address this challenge. Our study reports from an ongoing effort to deploy an integrated system based on Microsoft SharePoint (MSP) technologies[1]. We trace out local enactment (e.g. workarounds) of MSP, but more importantly demonstrate the cross-contextual nature of this enactment i.e. how workarounds in one context affect local appropriation of MSP in another context.

The structure of the remainder of this paper is organized as follows. In the next section we conceptualize the use of integrated information systems. Then, we outline our research approach and introduce historical context and intentions of changing collaborative infrastructure in OGC. Thereafter, we illustrate and discuss how local workarounds, as a result of the tight integration in MSP, shape use patterns across contexts. Finally, we provide analytical implications for studying the use of integrated IS and offer practical implications for managing generic infrastructures.

---

[1] http://www.microsoft.com/SharePoint/default.mspx

# 2 CONCEPTUALISING THE USE OF INTEGRATED SYSTEMS

## 2.1 Using technology in a situated context

Mirroring a more general interest in the social sciences for practice theory (Gherardi 2000; Savigny, Knorr-Cetina et al. 2001), information systems research has for some years studied its principal 'practice' viz. the practices that go into the use of information systems. Analyses of how users perceive, appropriate and subsequently use information systems demonstrate the highly contextual or situated nature of the use (or practice of use, if you want) of technology.

An early and influential contribution was Gasser's (1986) study of users' strategies of fitting, augmenting and working around the intentions inscribed into the functionality of the system. Gasser (ibid.) empirically illustrated what people actually do when confronted with rigid and unreliable computing procedures. The author vividly illustrated that users in fact do not use information systems as they are designed, but invent various ad-hoc strategies to fit the technology for a particular task. The major conceptualization from this study was the notion of *workaround*, which refers to "using computing in ways for which it was not designed or avoiding its use and relying on an alternative means of accomplishing work" (ibid., p.216).

While Gasser (1986) studied the rigid inventory control system, other more flexible types of systems were studied as well. For instance Orlikowski (1996) investigated how quite small customer support department used a new system to provide a better service for customers. Orlikowski (ibid.) illustrated how users contingently appropriate technology over the time. The central characteristic of appropriation is continuous change with unpredictable trajectory ('improvisation') rather than stability.

Central to these studies is to understand how technology is used in a situated context. While the notion of context is certainly vague (Chalmers 2004), context, in this case, is limited to individual actor's engagement and "recurrent interaction with the technology at hand" (Orlikowski 2000, p. 47).

## 2.2 Integrated systems

Several scholars have been interested in how integrated systems are used ('enacted'). For instance, Boudreau et al. (2005) in their recent study showed how, despite inherent rigidity of an ERP system, users are working-around the system in unintended ways. They argue that users first avoid the system (due to inertia), later learn by improvising (rather than in formal training) and finally reinvent the system in not-planned ways. Thus, the authors emphasize the human agency perspective over the technological logic and argue that "technology's consequences for organizations are enacted in use rather than embedded in technical features" (ibid. p.14). Other researchers have similarly emphasized the impossibility of large-scale systems to be universal across contexts due to local relevance or cultural fit (Joshi, Barrett et al. 2007). Some researchers have suggested that workaround is an intrinsic part of every day work rather than negative or unwanted effect (Rolland and Monteiro 2002).

In general, it was suggested that technologies would always drift from initial plans due to the improvisational capability of a human actor. In turn, the same technology can produce contrasting effects in similar organizational contexts (Robey and Sahay 1996) and these should be addressed employing the logic of opposition (Robey and Boudreau 1999).

Overall, studies on IS use tend to overemphasize local practices and do not "adequately address the longer-term co-evolution of artefacts and their social settings of use" (Pollock, Williams et al. 2007, p.257). Certainly, the relationship between contexts and spanning-effects are discussed by several scholars (Hanseth, Ciborra et al. 2001; Ellingsen and Monteiro 2006), however, there are few if any studies which conceptualize how local workaround does influence other contexts of use. For instance Boudreau et al. (2005) do identify the relationship between local appropriations: "An error occurring at one level of the system would have a ripple effect at other levels" (ibid., p.13), however they do not elaborate on the issue, nor do the authors elaborate what is the role of local enactments in larger contexts: "one cannot categorically argue that unintended actions are good, any more than one can argue that they are bad. Like any other aspect of organizational behaviour, evaluations of effectiveness are relative, not absolute" (ibid., p.16).

To sum it up, the study of integrated information systems has been framed to date largely along the lines of practice theory in the sense that local strategies for appropriation and use have been highlighted. Our study supplements this with a more systematic attention to the interdependence of cross-contextual appropriation mediated by integrated systems.

## 3    METHOD

We report from an ongoing longitudinal research project started in January 2007. Our research approach can be conceptualized as an interpretive case study (Walsham 1993) as we "attempt to understand phenomena through the meanings that people assign to them" (Klein and Myers 1999, p.69).

Data collection activities started at the beginning of 2007 with the primary aim to explore the change associated with the implementation of MS SharePoint technologies. We have employed 3 modes of data gathering: informal and formal interviews, observation and document studies.

We have conducted 25 in-depth interviews, on average lasting 1-1.5 hours. First interviews were open ended and aimed to identify IT strategic visions and implementation activities related to MS SharePoint. During later interviews, we analysed specific infrastructural components, work practices or individual engagements with technology. The technological complexity and intentions behind the new infrastructure were discussed with developers, administrators and managers of the collaborative infrastructure. The use of collaborative infrastructure was explored with actors from several organizational units. Interviewed users represent such disciplines as technology managers, human resources, senior researchers and various engineers involved in oil and gas production activities.

Participatory observations and informal discussions were mainly carried out in one of the OGC research centres, where both authors were granted access since the beginning of data collection. Since January 2008, one of the authors has been granted an office space, an access badge and access to OGC IT network. Since then, the researcher has been spending 2-3 working days a week in the research centre. Significant amount of time spent on-site forms the understanding of how work is carried out in practice and what problems and frustrations users experience on a daily basis. Additionally, being on-site gives an opportunity to have informal but informative chats around a coffee machine or during lunch breaks.

The third major empirical data source is internal OGC documents. We have extensively studied strategic documents related to planning and implementation activities of MSP. Additionally, we analysed technical descriptions, formal presentations and training materials on various MSP infrastructural components. A number of policy documents, which define how particular technology should be used or how specific work has to be carried out, were studied in detail. Finally, OGC intranet portal provided extensive contextual information on diverse OGC activities.

Data analysis is ongoing and iterative. Considering changing researchers involvement and overlapping but not the same research focus, the analysis of empirical data has many trajectories. This difference gives us a unique opportunity to analyse implementation process from slightly different perspectives. It is quite often that after interviews, if conducted together, we have a discussion and analyse what new aspects we have uncovered or what needs more attention in the subsequent data collection steps. In our faculty, there are several actors (not only the authors of this paper) exploring MSP implementation activities in OGC. We meet and discuss quite often either around a coffee machine or having more formal discussion sessions. Considering that the authors of the paper are involved researchers, significant part of data analysis and validation is actually occurring with the help of OGC actors. During informal or formal meetings, we frequently present our findings to various OGC actors. In turn, we are challenged, supported or directed to the issues that need more attention. For instance, several record's information managers supported our early findings on the metadata use in research and development activities, but we received extensive comments and suggestions to collect more empirical data in operative environments. Adjustments to some generalizations were made and empirical data collection directions were embraced.

In general, empirical data is classified in broad themes reflecting specific organizational project, practice or technical component. Such classification is neither all encompassing nor exhaustive; it is rather overlapping and continually changing. Theory has an important role in the analysis process. It provides an analytical lens to sort out and reclassify empirical data. For instance, in relation to this paper, the concept of workaround implied to determine when is a workaround (in relation to formal policy) and classify empirical data according to why, where and how workarounds are practiced. The concept of 'generification' implied to analyse how local workarounds 'travel' across contexts and what effects they produce.

# 4    CASE: COLLABORATION AND INTEGRATION

## 4.1    Context and history

Established only in the 1970s, the global oil and gas company (OGC, a pseudonym) has grown from a small, regional operator in Northern Europe to a significant energy company, currently employing some 30.000 people with activities in about 40 countries across 4 continents. OGC has grown largely organically, but with selected, important national and international acquisitions. Facing limited growth potential in its region of origin, OGC is actively pursuing a strategy to grow globally. To boost its financial capacity and flexibility, in the 1990s OGC diversified and expanded its shareholder ownership including getting listed at the New York Stock Exchange.

Alongside its growth in size, geography and business areas, OGC has been engaged in a number of corporate-wide initiatives to improve communication and collaboration. These initiatives have relied heavily on information systems. The first comprehensive effort to establish a corporate, collaborative information systems infrastructure was in the early 1990s (Monteiro and Hepsø 2002), at a time of oil industry recession, falling oil prices and dollar rates. Centralization, standardization and market orientation of IT services was the direct outcome of several projects whose primary aim was to solve the problems of fragmented and incompatible IT solutions. The outcome of standardization activities led to the establishment of the Lotus Notes-based collaborative infrastructure.

The Lotus Notes based infrastructure has proven successful inasmuch as it has been widely used for a range of different purposes. A key vehicle for facilitating collaboration within projects in OGC has been Lotus Notes Arena (Arena for short) databases for collective storing and dissemination of documents. The challenge, however, with the Lotus Notes based infrastructure has been to promote communication *across* the project-defined boundaries of the Arena databases. The Arena databases had no central indexing functionality, meaning that it was impossible to retrieve a document by searching if one did not know which database to search. With Arena databases thriving apparently 'out of control' – there were some 5000 databases by the latest estimates – locating relevant information stored outside your immediate project scope was non-trivial. Each user had in addition access to both personal (G disc) and departmental storage (F disc) areas. In short, information was scattered and duplicated over many local storage arenas.

### 4.2 New collaborative strategy – higher efficiency with tighter integration

To overcome the problems with Lotus Notes and establish more effective ways of collaboration, coordination and experience transfer, OGC formulated a new strategy in 2001. According to this strategy, OGC already had a set of general collaboration tools, but "these tools are poorly integrated", and "there is a particular need for better and more integrated coordination tools, better search functionality and improved possibilities for sharing information with external partners" (internal strategy documents). The change in the collaborative infrastructure was defined as a necessity and catalyst in order to achieve goals formulated in the strategy. The decision was made in 2003 and the rollout of a new infrastructure based on the Microsoft SharePoint (MSP) started. MSP was selected exactly for its potential to overcome the fragmentation resulting from project-specific Arena databases. Recent accounting regulation in the aftermath of Enron added pressure to ensure more systematic and consistent documentation of business decisions to inform the stock market and the public.

MSP is a core element in the new OGC collaborative infrastructure. The central element of MSP is so-called team site (TS), the virtual arena for collaboration. TS provides functionality to check-in and check-out documents, post announcements, share links and create discussion boards. Another important element of TS is a so-called workspace. A workspace is a web site connected to a TS (sometimes called baby-team site), used for production and sharing of a specific document or collection of documents. While MSP is mainly used for documents management, the technology is integrated with a corporate-wide search engine, an archive system and MS Exchange system.

While the technology itself (MSP) is customizable for specific contexts, the OGC decided to make the solution as generic as possible so that it would fit all contexts (internally it is referred as one-size-fits-all strategy). The strategic choice to rollout 'out-of-the-box' solution with minimum customization was highly influenced by the previous implementation experiences. In particular, the straightforward MSP implementation process was planned in contrast to recent experiences with an opposite (extensive customization) strategy when implementing a several hundred million dollars worth corporate-wide ERP solution. These experiences were translated in the standardization of both the functionality and the interface of every TS. The only element that differentiates team sites is metadata. The metadata standard provides a common and standardized classification scheme on how the information should be classified. Thus, the metadata can be seen as the main element in the collaborative infrastructure, which should fit a generic TS to a specific local situation. The metadata standard represents quite complex classification scheme with 13 different 'elements' and corresponding 'sub-elements'. In total there are more than 120 sub-elements in the metadata standard. Taking into account all sub-elements, the standard describes "identity, authenticity, content, structure context and essential management requirements of information objects" (OGC internal).

## 5  ANALYSIS: CONTEXTUAL AND CROSS-CONTEXTUAL USE

What studies of contextual use of information systems have convincingly established is the importance, indeed, necessity, of users' active appropriation of technology to local circumstances and concerns. In other words, local workarounds (or appropriation, tweaking, improvisation, drift etc.) are not anomalies or design shortcomings but *constitutive* elements of working technologies (Rolland and Monteiro 2002).

Zooming in on two extended illustrations from OGC's implementation of MSP (the use of workspaces and classification of documents), we first reiterate this point. More importantly, however, we go on to identify how these local workarounds – as a result of the tight integration in MSP – shape use patterns in other contexts of use. Table 1 summarizes this. One way, then, to formulate the gist of our analysis is to say that the local strategies of appropriation, the prerequisite of working information systems, are simultaneously non-local side-effects significantly influencing patterns of use in other contexts (i.e. cross-contextual).

|                          | **Local context: appropriation**                          | **Cross-contextual: side-effects**                                               |
|--------------------------|-----------------------------------------------------------|----------------------------------------------------------------------------------|
| The use of workspaces    | Reproducing folder structure from entrenched practices    | Invisibility of information due to the incorrect information and access rights management |
| Classification of documents | Overriding default values to fit local context         | Undermining the search engine and distributing additional work                   |
|                          | Incorrect, but convenient classification of sensitive information | Availability of sensitive information and suspension of the search engine for 5 months |

*Table 1.   A summary of local (contextual) and non-local (cross-contextual) enactments of integrated information for two examples within the MSP infrastructure.*

5.1     Workspaces: local appropriation as replicating existing practices

Prior to the implementation of MS SharePoint an important decision not to have folder structure in team sites was made. This decision was made due to the technical inability to create a complex folder structure in team sites (limitations of the URL length). On the other hand, users in OGC had quite long experience with folder structure in Lotus Notes infrastructure and indeed as Boudreau et al. (2005) explain did have difficulties to 'forget' previous practices. In the initial stages of MS SharePoint implementation users tended to avoid the new system and used file servers to share information instead. As one manager explained, the amount of documents in file servers exploded when team sites were introduced. However, after some time users got acknowledged with the system and found out that it is possible to replicate previously existed folder structure with the help of workspaces:

> "From the beginning it was very clearly communicated that we are not allowed to use workspaces to replicate folder structure. That was the intention… however people have been using computers here for the last 15 years, so they actually continue to make folders with workspaces despite the fact that they are told no to do so… All our team sites have a pile of workspaces [the user navigates to one of the team sites and shows a list of approximately 30 workspaces]" (User working with operational support for offshore activities)

Overall, replicating folder structure with workspaces is quite popular workaround in OGC. In particular, the workaround is practiced in information-rich contexts. The 'popularity' of this workaround illustrates that users do not adhere to OGC policies but are actively engaging and experimenting with technology. The existence of such workaround is not surprising; it can be explained with the concept of 'installed base', used by Boudreau and Robey (2005). Such treatment of workaround requires shifting the focus from identification to explanation of the effects a workaround is producing. Precisely because the MSP infrastructure is integrated, local appropriations, outlined above, are not only local; they have cross-contextual implications.

The initial intention of using workspaces was related to the possibility for team site users to create 'areas' (i.e. workspaces) with custom access rights. In that sense, a workspace was considered as a temporary 'arena' in order to limit or expand the original access rights in a team site. However, in practice, users sometimes create folder-like workspace structures and use the functionality of limiting the access. Importantly, in its current configuration, corporate-wide search engine is only returning those documents that a user has access to. Thus, documents stored in workspaces with limited access rights will be visible and retrievable only to specified users and 'invisible' to others.

Cross-contextual effects are especially experienced by users working across contexts i.e. whose ability to find information depends not only on their skills, but also on how others manage information locally. For instance, a well engineer responsible for conducting well interventions across different fields explains:

> "It is quite often that we do not have access to necessary information. When planning a well intervention we have to know a lot of technical information about a particular well and history of the well in general [this includes information about previous challenges/problems, and conducted interventions with corresponding experience reports produced after each intervention]. Sometimes you do not find information just because you do not have access… so you have to call various people and ask… it is very time consuming and I know some people do not bother spending all their time on that… however, not having important information means more uncertainty during operation, and this can increase the risk and cost of operation." (Well engineer; emphasis added)

Oil and gas exploration, production and export activities span across many disciplines not only in the OGC, but a significant part of activities is carried out by various external contractors. For instance, while a plan to drill a new well is primarily produced by several internal disciplines, an external contractor can perform drilling. This means that for a certain period of time an external contractor needs access to information related to the new drilling activities and probably to some historical reports. It adds complexity to access management, and workarounds made some time ago tend to pop-up here:

> "It is quite often that I get a call asking for help to find information or to give access. So I have to use a lot of my time on this… I would like them [contractors] to be more independent … to avoid this [access problems], for instance, after a meeting with contractors I am sending two emails, one to internal OGC employees with a link to a document and another one to external ones with attachment." (Drilling engineer; emphasis added)

Many users are aware of cross-contextual effects of incorrect access rights management by explaining that they do not know whether a specific information is existing or not: "the worst thing is that if you don't find information it does not necessarily mean that information is non-existing" (Engineer). Such effects lead to uncertainty and distrust the capabilities of the search engine. In general, these examples show that local workarounds change use patterns across contexts. In that sense, working in an 'integrated environment' means shifting the focus from how the system fits locally to how the system fits across contexts.

5.2     Classification of documents: the power of the default value

The decision to have flat document structure did not fuel too much enthusiasm for end-users and, as argued in the previous section, invoked workarounds. By removing folder structure the implementation team understood that some alternative way to classify information should be provided. In turn, it was decided to develop a common predefined classification scheme, which would form the basis for information structure and help both to sort and retrieve information.

As we have argued in the previous section that thinking 'flat' (no folders) introduces some problems, common classification scheme also did provide challenges for end users. The notion of 'common' does not imply fits-all or having the same meaning across contexts, on the contrary, as Star (1991, p.44) explains, "no networks are stabilized or standardized for everyone". Thus, in some contexts classification is not acknowledged:

> "this metadata, it is bad… very often when you will store a document none of the provided values fit. For instance for this document I can choose from 10-12 different values… but they all do not fit… for this document I can select such values as 'none', 'agenda', 'minutes', 'presentation'… this document is presentation so 'presentation' value fits very well [the respondent starts laughing]" (Engineer working with oil and gas production; emphasis added)

While users are quite often confronted with metadata that does not fit in their contexts, the questions remain how they cope with this situation. Sometimes users just tend to ignore the existence of metadata and use such values as 'none' or 'miscellaneous'. It was planned to have a controlled vocabulary (the values are predefined in advanced) and all policies state that users should use provided metadata. Interestingly enough, MSP functionality allows to delete provided metadata values and create new ones. In turn, the second enactment strategy is to develop new values that would make more sense in local context:

> "We have replaced provided metadata values with the new ones, which actually represent the activities we are working on in the project. Provided values were meaningless in relation to this project, so it would make no sense to use them" (User working in R&D)

Another challenging issue related to classification in OGC is the classification of sensitive information. Users are provided with rather simple security classification scheme to identify which documents can be available to anyone (open), to all OGC employees (internal), to specific groups (restricted distribution) or to selected individuals (confidential). Security classification is managed in team site and on the workspace level, implying that confidential document should be placed in a confidential team site rather than in an open one. The security classification scheme can be described as simple and intuitive, however, in practice, the definition of what is 'confidential' and how it should be handled (for the sake of convenience or additional work) is not given.

In turn, local enactments diverge from formal policies. It was, up to now, quite a 'standard' to store personal information on the private team site, which is not classified as confidential. In some instances, due to convenience reasons, a classified report from an external company was stored in an 'open' projects team site, rather than in a workspace with restricted access. Perhaps the most 'problematic' enactment of security classification was to store, due to unawareness and additional work, human resources related information in not confidential team sites.

Overall, classification, being an inseparable part of everyday work, is not given but has to be enacted in practice (Bowker and Star 1999). The problematic aspect, as we have illustrated above, is that imposed common classification has local variations (workarounds). While the problem of global standardization and local variation has received quite some attention (Star 1991), it is much less clear how local variations 'travel' across contexts. Essentially, common (shared and used across contexts) classification can be characterized as inherently having cross-contextual aspects. Such aspects are captured with writing/reading metaphor: "any reading and writing artefact that accumulates inscriptions cannot but coordinate [constrain and transform] the activities which write and which read these inscriptions" (Berg 1999, p.391). Thus, working-around common classification not only erodes the common and controlled character of classification, but also automatically imposes a certain amount of additional work for actors across contexts.

Such cross-contextual effects can be nicely illustrated with an example of planning and drilling a new well. While planning activities are conducted onshore, drilling process is to a large extent managed offshore. Planning is a collective effort of various disciplines and results in producing several documents. Central documents describe the whole drilling program, detailed drilling procedures, possible risks and a checklist, just to mention some. Drilling a new well means producing at least some 200 documents, which have to be stored in team site(s). Since they are stored in a flat structure, metadata is the primary sorting and filtering mechanisms. The problem according to one engineer is that "while the metadata values are not very bad, people sometimes do not use them or use them wrong". In turn, incorrect use of metadata in one context, produce effects in other contexts:

> "sometimes I get a call in the evening from offshore people saying that they have been searching for a specific document for an hour or so with no success… to avoid this we have developed a practice [which is unofficial i.e. a workaround] that for every new drilling program, a drilling engineer [working onshore] creates an excel document containing links to documents that are most important for drilling engineers working offshore. It is additional work as we [engineers working onshore] have to update those excel documents during drilling, but then offshore people have much better overview." (Drilling engineer working onshore; emphasis added).

In that sense local workaround (wrong classification of documents) produce additional work in other contexts and later trigger other workarounds (engineers developing an excel sheet with links). Thus, local workarounds can produce ripple effects. More importantly, achieving working infrastructure entails collective enactment across contexts rather than local enactment.

Another, more substantial cross-contextual effect, was enacted with security classification. Documents were not always classified as intended. Incorrect classification of documents, due to the tight integration with the corporate search engine, made possible for confidential information to be available to many more than it should be. For instance, during an interview with two system administrators responsible for technical aspects of MS SharePoint infrastructure, we were shown the possibilities of the search engine. One administrator entered the name of his colleague sitting besides to demonstrate how the search engine works in practice. Among the first results, a document containing the administrator's work evaluation appeared. It was an embarrassing moment, for administrators in particular. Such document contains personal information, and it should be available only to few persons and certainly should have not been retrieved in that situation. Similar local enactments not only propagated across contexts but accumulated as well. The situation, according to one manager "got out of control, since too much sensitive information due to incorrect classification was available for way too many users". Since it was not impossible to apply any 'quick fix', the search engine was suspended. The corporate-wide search service was not available to users for 5 months, the period that was used to 'clean-up' incorrect classification and develop the approach that would prohibit such incorrect enactments later. Compulsory training sessions, technical usability improvements, control routines and other initiatives are currently executed to prevent such effects. However, as it is now acknowledged in some management levels, order without workarounds is out of reach: "we may hope for altered attitudes and more care taken in the future – however, all the time search has been suspended, people have been working as before (but all the errors have been "invisible" as search was not available)" (a recent presentation on information security in OGC).

# 6    IMPLICATIONS AND CONCLUSIONS

We draw two sets of implications from our study of the deployment of a MS SharePoint based information infrastructure in OGC: one analytic and one practical.

Analytically, our study demonstrates the rich array of strategies and improvisational acts that go into the local appropriation of technology. Existing research (see above) has vividly illustrated process of social shaping of technology, i.e. how both flexible and rigid technologies are shaped in situated contexts. In line with the 'practice perspective' of Orlikowski (2000), our findings confirm that local workarounds, tinkering and 'situated improvisations' are not anomalies or design shortcomings but constitutive elements of working technologies.

More importantly, however, our study continues to address the nature of non-local – what we in this paper have dubbed cross-contextual – effects that are embedded in the appropriation of integrated systems. For sure, the mere existence of cross-contextual effects has been acknowledged before. Boudreau and Robey (2005) for instance point out that local practices may have 'ripple effect' beyond the local context. Yet they fail to develop this observation into a more systematic framework or make it subject to substantial theorising (see Hanseth et al. (2006) for an exception). One implication of our study, then, is to contribute to a higher visibility of cross-contextual effects of the use of integrated information systems. Systematic attention to cross-contextual side-effects extend the way the use of integrated information systems has been conceptualized to date. As opposed to largely local independent contexts of enacted technology, the use of integrated systems implies the interdependent enactment across contexts now linked as a result of integration. This entails considering the technological, and in particular integrative technological, detail more seriously than the previously outlined 'practice lens' (Orlikowski 2000) or the application of the 'practice lens' in integrative environments (Boudreau et al. 2005). In this way, our study is addressing the existing overemphasis on human agency and contributes to the studies on mutual shaping of information technology and its use, to approach equally important question on how technologies constrain the trajectory of enactment (Kallinikos 2004; Doherty, Coombs et al. 2006). The empirical illustration above of how the accumulation of multiple, local appropriations added up to the effect of closing down the corporate-wide search engine service for almost half a year is difficult to arrive at with attention largely focused on local, contextual or situated 'enactment' of technology.

Practical implications of our study relate to the (project) management of embarking on large-scale, comprehensive infrastructure like efforts of the type reported here. At the core, the insight that due to the scale and heterogeneity of work practices and existing technological interdependencies, workarounds need to be considered as constitutive elements of working infrastructures rather than anomalies, design shortcomings or unexpected effects. More importantly, working-around integrated systems cannot be any longer considered as local phenomena, which could be black-boxed and ascribed to a specific context. Prototyping early versions of integrative technology, for instance, has more limited value than for non-integrated technologies. While the attractiveness of integrated technology is based on seamless cross-contextual information exchange, such technological platform also comes (quite often as a surprise) with inherent cross-contextual effects. Simply put integrative technology is more than often evaluated from a perspective of what positive effects it can bring, while underestimating how local (small and perhaps unimportant at first) activities can produce great effects some time later across contexts. Essentially, the benefits of integrated systems can only be realized if they are fitted across contexts rather than in some local contexts.

The second practical implication relates to the evaluation of cost and benefits (Goodhue, Wybo et al. 1992) of generic, corporate-wide integrative infrastructures. We consider the term evaluation from an interpretive perspective (Walsham 1993, p.165-186) rather than an economic one. 'Costs', include both, the developer's effort to establish technological platform and users adjustments of technology to his/her needs. Generic solutions are adjustable and can fit quite well in some, but most often not in all contexts. If generic technology does not fit, additional work ('costs') has to be carried out. Classification is an illustrative example of this. We have exemplified how locally irrelevant classification will require additional users work to make it meaningful locally. The same happens with generic technological functionality. For instance, some users in OGC, unsatisfied with MSP functionality, voluntarily and not in accordance with existing policies, invest their time in implementing, learning and using more flexible technologies (such as Groove) or social software solutions (Wiki's). One way to evaluate then is to consider, by percentage, to how many actors the generic technology does fit. A qualitative alternative is to consider whether generic technology fits well in specific (not excluding core) business activities. In our evaluation, we are employing the latter perspective, and have illustrated throughout the paper that both development and use costs are high, and we doubt whether they outweigh the benefits. As the IS management literature suggests, the implementation of new technology should be cost-efficient. In turn, the management should take into account not only the costs of establishing integrated technical infrastructure, but the invisible (i.e. workarounds) costs as well. In that sense, it should be made explicit who, how much and when will pay the costs of having the generic integrated infrastructure.

# REFERENCES

Avgerou, C. and C. Ciborra (2004). The Social Study of Information and Communication Technology, Oxford University Press

Azad, B. and N. King (2008). "Enacting computer workaround practices within a medication dispensing system." European Journal of Information Systems 17(3): 264.

Berg, M. (1999). "Accumulating and Coordinating: Occasions for Information Technologies in Medical Work." Comput. Supported Coop. Work 8(4): 373-401.

Boudreau, M.-C. and D. Robey (2005). "Enacting Integrated Information Technology: A Human Agency Perspective." Organization Science 16(1): 3.
Bowker, G. C. and S. L. Star (1999). Sorting things out classification and its consequences. Cambridge, Mass., MIT Press.
Chalmers, M. (2004). "A Historical View of Context." Computer Supported Cooperative Work (CSCW) 13(3-4): 223 - 247.
Doherty, N. F., C. R. Coombs and J. Loan-Clarke (2006). "A re-conceptualization of the interpretive flexibility of information technologies: redressing the balance between the social and the technical." European Journal of Information Systems 15(6): 569.
Ellingsen, G. and E. Monteiro (2006). "Seamless Integration: Standardisation across Multiple Local Settings." Computer Supported Cooperative Work V15(5): 443-466.
Gasser, L. (1986). "The Integration of Computing and Routine Work." ACM Transactions on office Information systems.
Gherardi, S. (2000). "Practice-Based Theorizing on Learning and Knowing in Organizations." Organization 7(2): 211-223.
Goodhue, D. L., M. D. Wybo and L. J. Kirsch (1992). "The Impact of Data Integration on the Costs and Benefits of Information Systems." MIS Quarterly 16(3): 293.
Hanseth, O., C. U. Ciborra and K. Braa (2001). "The Control Devolution ERP and the Side-effects of Globalization." The Data base for advances in information systems 32(4): 34-46.
Hanseth, O., E. Jacucci, M. Grisot and M. Aanestad (2006). "Reflexive Standardization: Side Effects and Complexity in Standard Making." MIS Quarterly 30: 563.
Joshi, S., M. Barrett and G. Walsham (2007). "Balancing local knowledge within global organisations through computer-based systems: an activity theory approach." Journal of Global Information Management 15(3): 1-19.
Kallinikos, J. (2004). Farewell to Constructivism: Technology and Context-Embedded Action. The Social study of information and communication technology : innovation, actors, and contexts. F. Land, C. Avgerou and C. U. Ciborra. Oxford, Oxford University Press**:** 140-161.
Klein, H. K. and M. D. Myers (1999). "A Set of Principles for Conducting and Evaluating Interpretive Field Studies in Information Systems." MIS Quarterly 23(1): 67.
Monteiro, E. and V. Hepsø (2002). "Purity and Danger of an Information Infrastructure." Systemic Practice and Action Research 15(2): 145-167.
Orlikowski, W. J. (1996). "Improvising Organizational Transformation Over Time: A Situated Change Perspective." Information Systems Research 7(1): 63.
Orlikowski, W. J. (2000). "Using technology and constituting structures: A practice lens for studying technology in organizations." Organization Science 11(4): 404.
Pollock, N., R. Williams and L. D'Adderio (2007). "Global software and its provenance: Generification work in the production of organizational software packages." Social Studies of Science 37(2): 254-280.
Robey, D. and M.-C. Boudreau (1999). "Accounting for the Contradictory Organizational Consequences of Information Technology: Theoretical Directions and Methodological Implications." Information Systems Research 10(2).
Robey, D. and S. Sahay (1996). "Transforming work through information technology: A comparative case study of geographic information systems in county government." ISR 7(1).
Rolland, K. H. and E. Monteiro (2002). "Balancing the local and the global in infrastructural information systems." Information Society 18(2): 87.
Savigny, E. v., K. Knorr-Cetina and T. R. Schatzki (2001). The Practice turn in contemporary theory, Routledge.
Star, S. L. (1991). Power, technologies and the phenomenology of conventions: on being allergic to onions. A Sociology of monsters essays on power, technology and domination. J. Law. London, Routledge**:** 26-56.
Walsham, G. (1993). Interpreting information systems in organizations. Chichester, Wiley.
Walsham, G. (2001). Making a world of difference: IT in a global context. Chichester, Wiley.